\documentclass[prd,preprint,showpacs,groupedaddress]{revtex4-1}
\usepackage{amsmath}
\usepackage{amsfonts}
\usepackage{amssymb}
\usepackage{geometry}
\usepackage{natbib}
\usepackage[english]{babel}
\usepackage{graphicx}
\usepackage{epstopdf}
\usepackage{subfigure}
\usepackage{indentfirst}
\usepackage{mathrsfs}

\begin{document}

  \setlength{\parindent}{2em}
  \title{Joule-Thomson expansion of the Bardeen-AdS black holes}
  \author{Cong Li} \author{Pengzhang He} \author{Ping Li} \author{Jian-Bo Deng} \email[Jian-Bo Deng: ]{dengjb@lzu.edu.cn}
  \affiliation{Institute of Theoretical Physics, LanZhou University, Lanzhou 730000, P. R. China}
  \date{\today}

  \begin{abstract}

  The Joule-Thomson expansion process is studied for Bardeen-AdS black holes in the extended phase space. Firstly, we get Joule-Thomson coefficient and find that the divergent point of Joule-Thomson coefficient coincides with the zero point of temperature. The inversion curves are also obtained from the zero point of Joule-Thomson coefficient. Then the minimum inversion temperature and the corresponding mass are obtained. In addition, the ratio between minimum inversion and critical temperature for Bardeen-AdS black holes is also calculated. We obtain the isenthalpic curve in $T$-$P$ graph and demonstrate the cooling-heating region by the inversion curve. An interesting phenomenon we get is that black hole whose mass to charge ratio is below the critical value is always in heating process. The same phenomenon can be also obtained from the charged AdS black holes.
  \end{abstract}



  \maketitle

  \section{Introduction}

  Since Bekenstein and Hawking firstly studied the properties of black hole thermo\-dynamics\cite{1,2,3}, thermodynamics of black holes has attracted lots of attention. Anti-de-Sitter (AdS) black hole thermodynamics has been first studied by Hawking and Page and Hawking-Page phase transition was found\cite{4}. What's more, properties of the charged AdS black hole thermodynamics were studied in \cite{5,6} where a van der Waals like phase transition was found in the charged AdS black holes.
\par
  In the extended phase space, where the cosmological constant $\Lambda$ is regarded as a thermodynamic variable related to the positive pressure $P$ in AdS space, the analogy between black holes and van der Waals liquid-gas system\cite{7} was studied and the critical exponents\cite{8} have also been found. This analogy has been generalized to different AdS black holes, such as Kerr-black holes, Gauss-Bonnet black holes, the higher dimensional black holes and the Lovelock gravity, etc.\cite{9,10,11,12,13,14,Kastor_2010}. It was shown in \cite{7} that considering the cosmological constant $\Lambda$ as pressure $P$, the black hole mass $M$ was the enthalpy $H$ rather than the internal energy $U$. Furthermore, the reentrant phase transition\cite{9,15} and the triple point\cite{11,16} for AdS black holes have also been studied in the extended phase space, proposing that AdS black holes behaved similarly to the ordinary thermodynamic systems.
\par
  Recently, {\"O}kc{\"u} and Ayd{\i}ner studied the Joule-Thomson expansion process in the charged AdS black holes with the analogy between charged AdS black holes and van der Waals system\cite{17}. For Joule-Thomson expansion in classical thermodynamics, gas at a high pressure passes through a porous plug to a section with a low pressure and during the process the enthalpy is unchanged. An interesting phenomenon during the Joule-Thomson expansion process is that the $T$-$P$ graph is divided into two parts, the cooling region and the heating region, by the inversion curve. The isenthalpy and inversion curves were first studied in charged AdS black holes and Kerr-AdS black holes\cite{18}. Then the works were generalized to AdS black holes with a global monopole\cite{19}, higher dimensional AdS black holes\cite{20}, AdS black holes in Lovelock gravity~\cite{21} and charged Gauss-Bonnet black holes in AdS space\cite{22}, etc. All the articles above show that the inversion curves $T_{i}$-$P_{i}$ only have positive slope, which was different from the inversion curves in van der Waals system including both positive and negative slopes. However, black holes in previous discussion of Joule-Thomson expansion all exist singularity. In addition, Bardeen-AdS black holes have no singularity inside the horizon, called regular black holes. So, we would like to to see whether the singularity have the affect on the Joule-Thomson expansion for studying Bardeen-AdS black holes.
\par
  This paper is organized as follows. In Sec.\uppercase\expandafter{\romannumeral2}, we briefly review Bardeen-AdS black holes. In Sec.\uppercase\expandafter{\romannumeral3}, we investigate Joule-Thomson expansion for Bardeen-AdS black holes. The isenthalpic and inversion curves have been obtained. Conclusions and discussion are given in Sec.\uppercase\expandafter{\romannumeral4}.

  \section{A brief review of Bardeen-AdS black holes }
  Bardeen black hole is known as a black hole without singularity\cite{23} and the corresponding action with $\Lambda$ term can be given by
  \begin{equation}
   \label{eq:S}
   S=\frac{1}{16\pi}\int d^4x\sqrt{-g}\left[R+\frac{6}{l^2}-4\mathcal{L}(F)\right],
  \end{equation}
  where $R$ is scalar curvature, $g$ is the determinant of the metric tensor, $l$ is AdS radius connected with $\Lambda$ through the relation $\Lambda=-\frac{3}{l^2}$. $\mathcal{L}(F)$ is a function of $F=\frac{1}{4}F_{\alpha\beta}F^{\alpha\beta}$ with $F_{\mu\nu}=2\nabla_{[\mu}A_{\nu]}$ given by\cite{24}
  \begin{equation}
  \mathcal{L}(F)=\frac{3}{2sq^2}\left(\frac{\sqrt{2q^2F}}{1+\sqrt{2q^2F}}\right)^{\frac{5}{2}}.
  \end{equation}
  The parameter $s$ in above equation is given by $\frac{|q|}{2M}$ where $q$ and M correspond to the magnetic charge and the mass of the black hole.
  The field equations derived from the action in Eq.~(\ref{eq:S}) are given by
  \begin{equation}
   G_{\alpha\beta}+\Lambda g_{\alpha\beta}=2\left(\frac{\partial\mathcal{L}(F)}{\partial F}F_{\alpha\lambda}F^{\lambda}_{\beta}-g_{\alpha\beta}\mathcal{L}(F)\right),
  \end{equation}
  \begin{equation}
   \nabla_{\alpha}\left(\frac{\partial\mathcal{L}(F)}{\partial F}F^{\beta\alpha}\right)=0,
  \end{equation}
  \begin{equation}
   \nabla_{\alpha}(*F^{\beta\alpha})=0.
  \end{equation}
  Then the line element of Bardeen-AdS black holes can be considered as
  \begin{equation}
   \label{eq:element}
    ds^2=-f(r)dt^2+\frac{1}{f(r)}dr^2+r^2(d\theta^2+sin^2{\theta}d\phi^2),
  \end{equation}
  where $f(r)=1-\frac{2m(r)}{r}$. Following the ansatz for Maxwell equation $F_{\mu\nu}=2\delta^{\theta}_{[\mu}\delta^{\phi}_{\nu]}Z(r,\theta)$~\cite{Eloy} and with the help of Eq.~(4), we can get
  \begin{equation}
    \label{F}
    F_{\mu\nu}=2\delta^{\theta}_{[\mu}\delta^{\phi}_{\nu]}g(r)sin\theta.
  \end{equation}
  Using the condition $dF=g'(r)sin\theta~dr\wedge d\theta \wedge d\phi=0$, we conclude that $g(r)=const.=g$. Hence, the field strength is $F_{\theta\phi}=-F_{\phi\theta}=qsin\theta$ with $F=\frac{g^2}{2r^4}$. Substituting the expression in Eq.~(2), we get
  \begin{equation}
    \label{L}
    \mathcal{L}(F)=\frac{3Mg^2}{(g^2+r^2)^{\frac{5}{2}}}.
  \end{equation}
  Moreover, the field equation (3) can yield a solution for $m(r)$ with
  \begin{equation}
    \label{m}
    m(r)=\frac{Mr^3}{(g^2+r^2)^\frac{3}{2}}-\frac{r^3}{2l^2}.
  \end{equation}
  So the metric of Bardeen-AdS black hole is
  \begin{equation}
    \label{f(r)}
    f(r)=1-\frac{2Mr^2}{(g^2+r^2)^\frac{3}{2}}+\frac{r^2}{l^2}.
  \end{equation}
  When $f(r_{+})=0$, we get
  \begin{equation}
   \label{eq:M}
    M=(1+\frac{r^2_{+}}{l^2})\frac{(r^2_{+}+q^2)^\frac{3}{2}}{2r^2_{+}},
  \end{equation}
  which satisfies the first law of black hole thermodynamics,
  \begin{equation}
   \label{eq:dM}
    dM=TdS+VdP+\Phi dq.
  \end{equation}
  Here the pressure $P$ in AdS black holes corresponds to the $\Lambda$ with
  \begin{equation}
   \label{eq:P}
    P=-\frac{\Lambda}{8\pi}=\frac{3}{8\pi l^2},
  \end{equation}
  and the conjugate quantity $V$ of the pressure $P$ can be given by
  \begin{equation}
   \label{eq:V}
    V=(\frac{\partial M}{\partial P})_{S,q}=\frac{4\pi r_{+}^3}{3}(1+\frac{q^2}{r_{+}^2})^\frac{3}{2}.
  \end{equation}
  The Hawking temperature $T$ can be obtained from the first derivative of $f(r)$ at the horizon\cite{25},
  \begin{equation}
   \label{eq:T}
    T=\frac{\kappa}{2\pi}=\frac{f'(r_{+})}{4\pi}=\frac{3r_{+}}{4\pi(r_{+}^2+q^2)}+\frac{3r_{+}^3}{4\pi(r_{+}^2+q^2)l^2}-\frac{1}{2\pi r_{+}}.
  \end{equation}
  Then, the equation of state $P=P(V,T)$ for Bardeen-AdS black holes can be obtained from Eqs.~(\ref{eq:P}) and~(\ref{eq:T}),
  \begin{equation}
   \label{eq:P1}
    P=\frac{4\pi r_{+}(q^2+r_{+}^2)T+2q^2-r_{+}^2}{8\pi r_{+}^4}.
  \end{equation}
  According to \cite{8}, the critical temperature can be calculated by
  \begin{equation}
   \label{eq:P_{c}}
   \frac{\partial P}{\partial v}=0,~~\frac{\partial^2 P}{\partial v^2}=0,
  \end{equation}
  where $v$ is $\frac{2r_{+}^3}{q^2+r_{+}^2}$ and we can get
  \begin{equation}
   \label{eq:T_{c}}
   T_{c}=\frac{7+\sqrt{273}}{\sqrt{2(15+\sqrt{273})}(21+\sqrt{273})\pi q}.
  \end{equation}

  \section{Joule-Thomson expansion }

  In this section, we will investigate Joule-Thomson expansion for Bardeen-AdS black holes. The main feature of this process is that enthalpy remains constant. In addition, the enthalpy $H$ can be identified as the mass $M$ of the AdS black holes\cite{7}. So, we can keep the black hole mass constant. Similar to the Joule-Thomson process with fixed particle number for van der Waals gases, we consider the canonical ensenmble with fixed magnetic charge $q$. Therefore, Joule-Thomson coefficient $\mu$, which characterizes the expansion, is given by ~\cite{22}
  \begin{equation}
   \label{eq:mu1}
   \mu={(\frac{\partial{T}}{\partial{P}})}_{M,q}=(\frac{\partial{T}}{\partial{r_{+}}})_{M,q}(\frac{\partial{r_{+}}}{\partial{P}})_{M,q}.
  \end{equation}
  From Eqs.~(\ref{eq:M}) and (\ref{eq:T}), the pressure and temperature written as a function of $M$ and $r_{+}$ are given
  \begin{equation}
   \label{eq:P(M,r)}
   P(M,r_{+})=\frac{-3(q^2+r_{+}^2)^\frac{3}{2}+6Mr_{+}^2}{8\pi r^2(q^2+r_{+}^2)^\frac{3}{2}},
  \end{equation}
  \begin{equation}
   \label{eq:T(M,r)}
   T(M,r_{+})=\frac{-(q^2+r_{+}^2)^\frac{5}{2}+3Mr_{+}^4}{2\pi r(q^2+r_{+}^2)^\frac{5}{2}}.
  \end{equation}
  So, one can obtain the Joule-Thomson coefficient
  \begin{equation}
  \begin{aligned}
   \label{eq:mu}
   \mu&=\frac{2r_{+}(9Mq^2r_{+}^4-6Mr_{+}^6+(q^2+r_{+}^2)^\frac{7}{2})}{3(q^2+r_{+}^2)^\frac{7}{2}(1-\frac{3Mr_{+}^4}{(q^2+r_{+}^2)\frac{5}{2}})}\\
      &=\frac{4q^4r_{+}+q^2(26r_{+}^3+48\pi Pr_{+}^5)-8(r_{+}^5+4\pi Pr_{+}^7)}{3(q^2+r_{+}^2)(2q^2-r_{+}^2-8\pi Pr_{+}^4)}.
  \end{aligned}
  \end{equation}
  Compare Eqs.~(\ref{eq:T}) and (\ref{eq:mu}), we can get that the divergent point of Joule-Thomson coefficient coincides with the zero point of temperature. The divergent point maybe reveals the information of Hawking temperature and corresponds to the extremal black hole.\\
  The zero point of Joule-Thomson coefficient is the point that distinguishes the cooling process and the heating process. From Eq.~(\ref{eq:mu}), one can obtain that the inversion pressure $P_{i}$ and the corresponding $r_{+}$ have the relation,
  \begin{equation}
   \label{eq:r1}
   2q^4+q^2(13r_{+}^2+24\pi P_{i}r_{+}^4)-4(r_{+}^4+4\pi P_{i}r_{+}^6)=0.
  \end{equation}
  The only positive and real solution of the above function is so long that we will not list them here. But we depict the relation between the inversion temperature $T_{i}$ and the inversion pressure $P_{i}$ in Fig. 1 with the help of Eqs.~(\ref{eq:T}) and~(\ref{eq:r1}). The inversion curves are not closed and there is only one inversion curve in comparison with van der Waals fluids. For low pressure, the inversion temperature decreases with magnetic charge $q$, whereas it increases with $q$ for high pressure.\\
  \begin{figure}[htp]
  \centering
  \includegraphics[width=10cm]{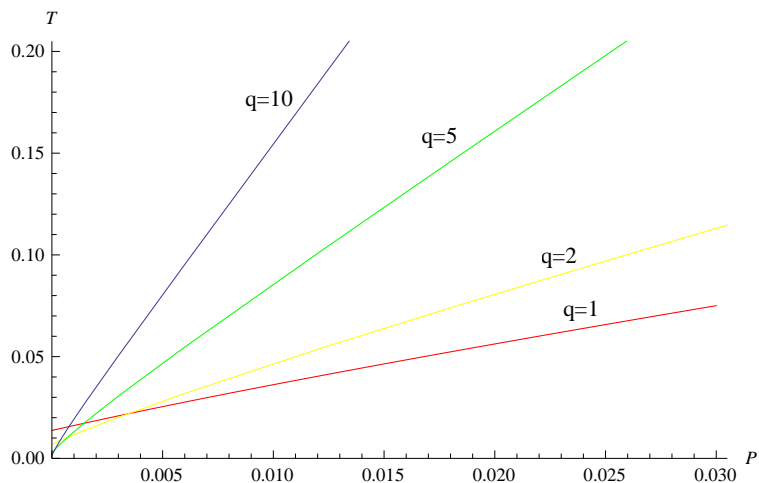}
  \caption{Inversion curves of Bardeen-AdS black holes in $T-P$ plane.}
  \end{figure}
  Then we would like to probe the ratio between the minimum inversion temperature $T_{i}^{min}$ and the critical temperature $T_{c}$ first. Note that the minimum temperature can be obtained by demanding $P_{i}=0$. The Eq.(\ref{eq:r1}) will be
  \begin{equation}
   2q^4+13q^2r_{+}^2-4r_{+}^4=0.
  \end{equation}
  The only physically meaningful root is given by
  \begin{equation}
   \label{eq:r_{+}}
   r_{+}=\frac{\sqrt{13+\sqrt{201}}q}{2\sqrt{2}}.
  \end{equation}
  One can substitute the root into Eq.~(\ref{eq:T}) and obtain the minimum inversion temperature,
  \begin{equation}
   T_{i}^{min}=\frac{-3+\sqrt{201}}{\sqrt{2(13+\sqrt{201})}(21+\sqrt{201})\pi q}.
  \end{equation}
  The ratio between minimum inversion and critical temperature is given by
  \begin{equation}
   \frac{T_{i}^{min}}{T_{c}}\approx0.545874.
  \end{equation}
  This ratio is not equal to the value in other black holes\cite{17,18,19,20,21}. This different ratio may be due to the non-linear electrodynamics field and the correctional volume of Bardeen-AdS black holes. Moreover, we can substitute Eq.~(\ref{eq:r_{+}}) into Eq.~(\ref{eq:M}) and obtain the mass corresponding to the minimum inversion temperature,\\
  \begin{equation}
   \label{eq:m_{min}}
   M_{min}=\frac{(21+\sqrt{201})^{3/2}}{4\sqrt{2}(13+\sqrt{201})}q\approx1.3571q.
  \end{equation}
  \begin{figure}[htbp]
  \centering
  \subfigure[]{
  \begin{minipage}{6.5cm}
  \centering
  \includegraphics[width=5.5cm]{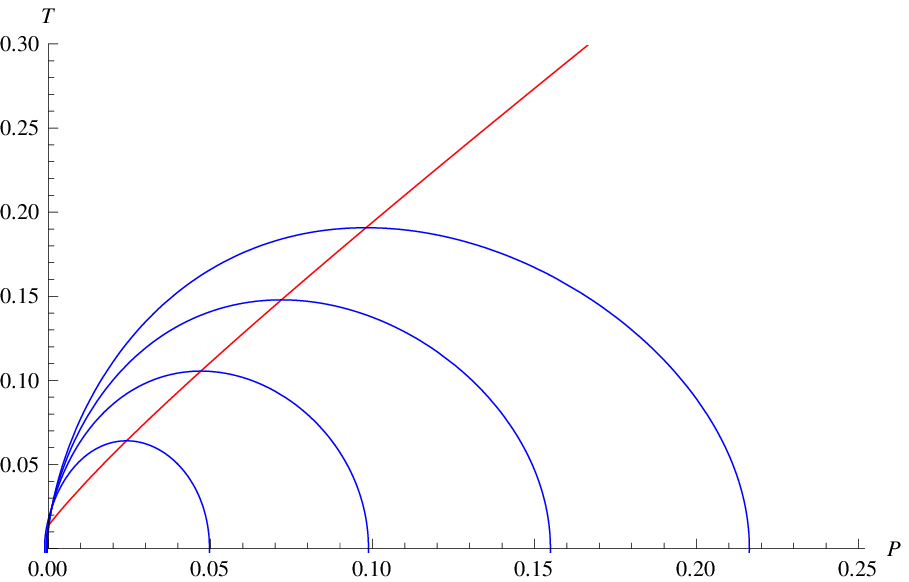}
  \end{minipage}%
  }%
  \subfigure[]{
  \begin{minipage}{6cm}
  \centering
  \includegraphics[width=5.5cm]{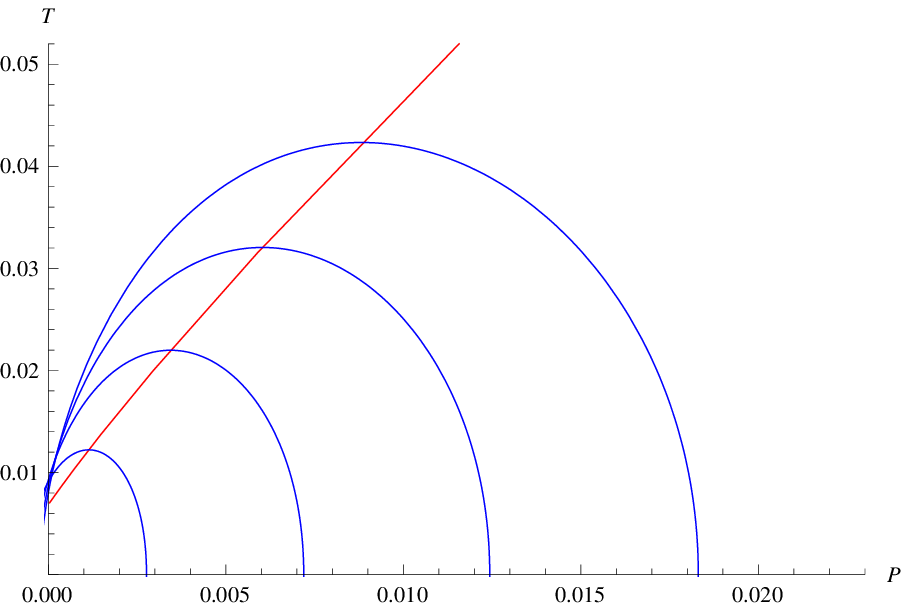}
  \end{minipage}
  }
  \subfigure[]{
  \begin{minipage}{6.5cm}
  \centering
  \includegraphics[width=6cm]{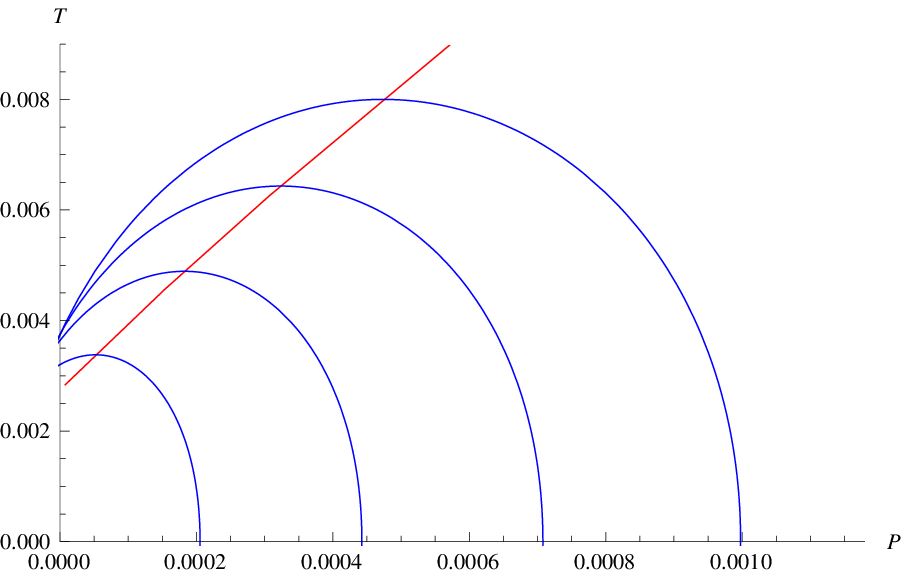}
  \end{minipage}%
  }%
  \subfigure[]{
  \begin{minipage}{6cm}
  \centering
  \includegraphics[width=5.5cm]{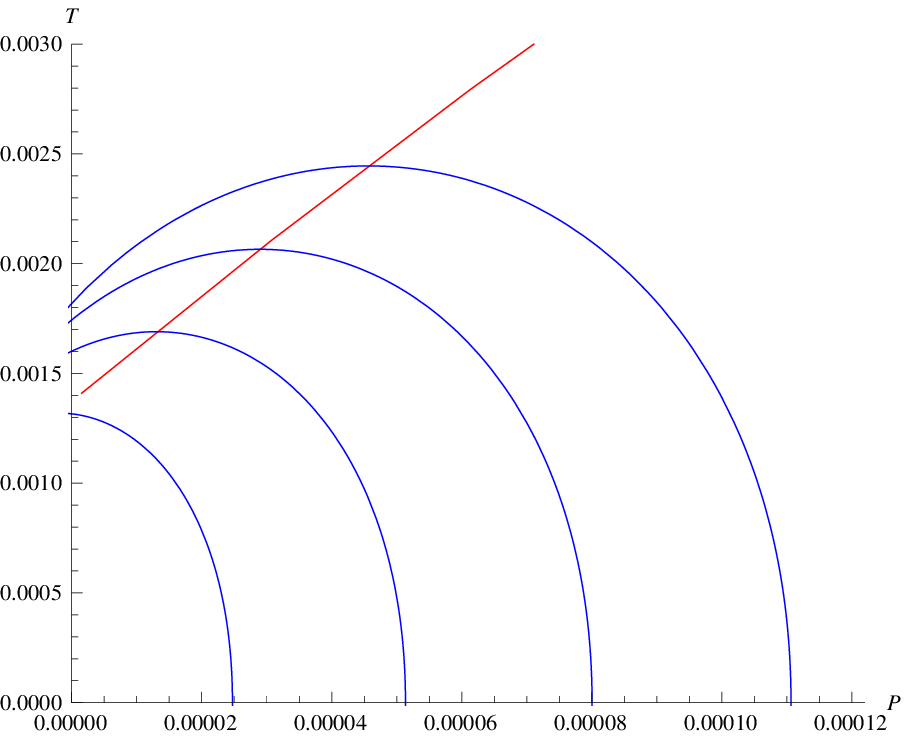}
  \end{minipage}
  }
  \caption{Inversion and isenthalpic curves of Bardeen-AdS black holes. The red lines present inversion curves and the blue ones are isenthalpic curves. From bottom to top, isenthalpic curves correspond to increasing values of $M$. (a) $q$=1 and $M$=2,~2.5,~3,~3.5. (b) $q$=2 and $M$=3,~3.5,~4,~4.5. (c) $q$=5 and $M$=7,~7.5,~8,~8.5. (d) $q$=10 and $M$=13.5,~14,~14.5,~15.}
  \end{figure}
  This corresponding mass will help us to determine the black hole whether have the inversion point in Joule-Thomson expansion. The isenthalpic and inversion curves for various values of magnetic charge $q$ has been drawn in Fig.~2. As can been seen from Fig.~2, the inversion curves divide the plane into two regions. The branch below the inversion curves corresponds to the heating process with negative slope. And the branch above the inversion curves corresponds to the cooling process with positive slope. It is interesting to see in Fig.~2 that one isenthalpic curve does not have the inversion point. Moreover, for $q=10$ in Eq.~(\ref{eq:m_{min}}), we have the mass $M_{min}\approx13.571$ which is greater than $M=13.5$. This also shows that the black hole with $q=10$ and $M=13.5$ does not have the inversion point. In other word, the black hole with $q=10$ and $M=13.5$ is always in heating process.\\
  \begin{figure}[htp]
  \centering
  \includegraphics[width=10cm]{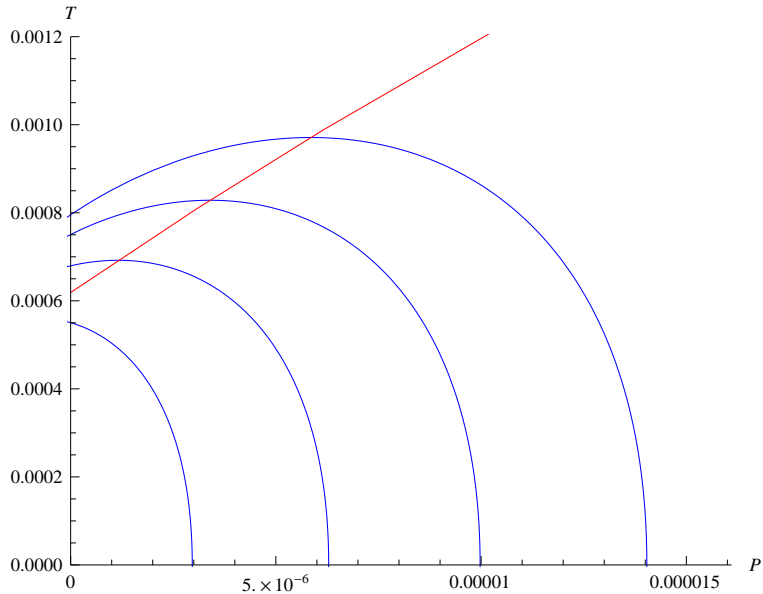}
  \caption{Inversion curves of charged AdS black holes in $T-P$ plane.~$Q$=35 and $M$=35.5,~36,~36.5,~37}
  \end{figure}
  The same phenomenon can alao be obtained from the charged AdS black holes with large $Q$ which is not given in \cite{17}. The figure we have plotted in Fig.~3. Similar steps can be used in charged AdS black holes\cite{17}, we get\\
  \begin{equation}
   \label{eq:charged M}
   M_{min}^{Q}=\frac{5\sqrt6}{12}Q\approx1.0206Q.
  \end{equation}
  Similarly, for $Q=35$ in Eq.~(\ref{eq:charged M}), we have the mass $M_{min}^{Q}\approx35.722>35.5$.\\
  \section{Conclusions and discussion}
  In this article, we studied Joule-Thomson expansion for Bardeen-AdS black holes in the extended phase space. We considered the expansion process with constant mass as the mass of AdS black holes was identified as enthalpy. Firstly, we got Joule-Thomson coefficient and found that the divergent point of Joule-Thomson coefficient coincided with the zero point of temperature. That is to say, the divergent point of Joule-Thomson coefficient reveals the information of Hawking temperature.
\par
  Then we plotted inversion curves with different magnetic charge $q$ and found there was only one inversion curve which was different from van der Waals fluids having two inversion curves. Moreover, we obtained the minimum inversion temperature $T_{i}$ and the corresponding mass. The ratio between minimum inversion and critical temperatures was calculated for Bardeen-AdS black holes. The ratio is not equal to the value which have been obtained in other articles and this different may be due to the non-linear electrodynamics field and the correction volume of Bardeen-AdS black holes.
\par
  We also plotted the isenthalpic and inversion curves in the $T-P$ plane. The inversion curve divides the $T-P$ plane into two regions. The region above the inversion curve is the cooling region with positive slope for isenthalpic curves. The region under the inversion curve is the heating region with negative slope for isenthalpic curves. An interesting phenomenon is that there is an isenthalpic curve does not have the critical point which means that the black hole is always in heating process. That is to say, as the pressure goes down, the temperature of horizon in this black hole can keep rising during the Joule-Thomson expansion. The mass of this black hole is also less than the mass corresponding the minimum inversion temperature. In addition, we also get the same phenomenon of charged AdS black holes with large charge $Q$. This phenomenon may be also obtained from other black holes.

\section*{Conflicts of Interest}
  The authors declare that there are no conflicts of interest regarding the publication of this paper.

\section*{Acknowledgments}
  We would like to thank the National Natural Science Foundation of China (Grant No.11571342) for supporting us on this work.

\section*{References}

\bibliographystyle{unsrt}
\bibliography{reference}

\end{document}